\documentclass[conference]{IEEEtran}
\IEEEoverridecommandlockouts
\usepackage{cite}
\usepackage{amsmath,amssymb,amsfonts}
\usepackage{algpseudocode}
\usepackage{algorithm}
\usepackage{graphicx}
\usepackage{textcomp}
\usepackage{xcolor}
\usepackage{url}
\def\BibTeX{{\rm B\kern-.05em{\sc i\kern-.025em b}\kern-.08em
    T\kern-.1667em\lower.7ex\hbox{E}\kern-.125emX}}
\begin{document}


\title{Peer-to-Peer Energy Trading meets IOTA: \\Toward a Scalable, Low-Cost, and \\Efficient Trading System}

\author{
\IEEEauthorblockN{Conor Mullaney, Adnan Aijaz, Nathan Sealey, Ben Holden}
\IEEEauthorblockA{
\text{Bristol Research and Innovation Laboratory, Toshiba Europe Ltd., Bristol, United Kingdom}\\
firstname.lastname@toshiba-bril.com}
}

\maketitle

\begin{abstract}
Peer-to-Peer (P2P) energy trading provides various benefits over conventional wholesale energy markets and makes renewable energy more accessible. This paper proposes a novel multi-layer P2P energy trading system for microgrids based on IOTA 2.0, which is a distributed ledger technology (DLT) primarily designed for Internet-of-Things (IoT) applications. The proposed energy trading system, which is a manifestation of a cyber-physical system (CPS), exploits the benefits brought by IOTA's unique ledger structure as well as the recently introduced IOTA smart contract protocol (ISCP). Further, it implements a uniform double-auction market mechanism and a hierarchical routing structure for interconnected microgrids. Performance evaluation demonstrates key benefits over wholesale markets as well as speed, energy efficiency and cost benefits over conventional blockchain-based P2P energy trading systems. 
\end{abstract}

\begin{IEEEkeywords}
Blockchain, CPS, DLT, IOTA, energy trading, microgrid, P2P, smart contracts, Tangle. 
\end{IEEEkeywords}

\section{Introduction}

Increased penetration of distributed energy resources (DERs) in the energy sector paves the way toward decentralization of current energy markets by allowing consumers to become prosumers \cite{ZHANG20181}. Typically energy generation based on DERs is intermittent and difficult to predict, requiring robust management systems to utilize them effectively. Peer-to-peer (P2P) energy trading is a promising approach for effective management of DERs in smart cities \cite{Arumugam2022}. The fundamental concept of P2P trading is that prosumers and consumers can buy and sell energy amongst themselves directly in a completely decentralized manner.  P2P energy trading promotes the use of renewable energy by empowering prosumers to actively participate in the energy market \cite{Tushar2020}. For example, a consumer household with solar panels can trade surplus energy with its neighbours. This will increasingly remove centralization in energy generation and diminishes the need for large non-renewable power plants. 

The key principles for designing P2P energy management systems (e.g., platforms) include decentralization, transparency, privacy, scalability, energy efficiency and user control. Decentralization benefits the entities in the system by removing their need to trust in a centralized third party. Transparency ensures that the system and all actors are acting honestly whilst ensuring privacy.  User control gives entities the choice as to what they share and with whom. Scalability becomes important due to rapidly growing demand; thanks to the growth in population and Internet-of-Things (IoT) devices. Energy efficiency of any system must be considered due to environmental concerns. 


Blockchain, which is the most well-known type of distributed ledger technology (DLT), has been identified as the key to facilitating P2P energy trading due to the inherent features of decentralization, resilience, privacy and security \cite{Thukral2021}. However, conventional blockchain technologies still face challenges when applied to a P2P energy trading use case. DLTs based on a directed acyclic graph (DAG) structure overcome many limitations associated with blockchains and have been investigated from the perspective of blockchain replacement in energy management systems \cite{mg}.


To this end, this paper advances state-of-the-art by developing a P2P energy trading system based on IOTA Tangle,\footnote{Our focus is strictly on the newer version, i.e., IOTA 2.0 (https://www.iota.org/) which is significantly different from the first version.} which is a DAG-based DLT specifically designed for IoT environments. The proposed energy trading system is a realization of cyber-physical systems (CPSs) approach and it provides a multi-layer trading functionality involving microgrids, IOTA smart contracts, and the Tangle. To the best of our knowledge, this is one of the first works investigating the use of IOTA Tangle and smart contracts for P2P energy trading.  
The main contributions of this work are summarized as follows:
\begin{itemize}
    \item We develop a P2P energy trading system that exploits the key benefits of \emph{IOTA Tangle} and \emph{IOTA smart contract protocol (ISCP) framework}. The proposed system provides key advantages of lower cost, lower transaction times, and high scalability compared to conventional systems.


    \item We implement a market mechanism via smart contracts employing a \emph{uniform double-auction} to match producers and consumers fairly in each bidding round, allowing microgrids to act as a single entity when interacting and trading. Our multi-layer trading system, underpinned by IOTA, ensures the transparency and security of transactions in a decentralized manner.

    \item We introduce a \emph{hierarchical routing structure} of interconnected microgrids to fulfil all excess supply/demand in the system per bidding round. This maximizes energy utility and adheres to demand response standards. 
    
    \item We provide testbed-based and simulations-based results to assess the proposed system in various trading scenarios using empirical data. Our results focus on both market-centric and DLT-centric evaluation.  
\end{itemize}


The rest of this paper is organized as follows. Section II covers related work and preliminaries on the subject matter. Section III describes our proposed IOTA-based P2P energy trading system. Section IV provides evaluation results and insights. The paper is concluded in section V with some directions for future work.


\section{Background and Preliminaries}
\subsection{Energy Trading Market and P2P Energy Exchange}
The problems and inefficiencies of wholesale centralized energy trading markets have been the focus of various studies.  The authors in \cite{Aitzhan2018} present security and single point-of-failure concerns. In \cite{Gregoratti2015}, it has been shown that centralization of power generation creates challenges for providing supply flexibility necessary to meet dynamic energy demands in an adequate time. Privacy \cite{Guan2019}, lack of competitive pricing \cite{Ettlin2018} and resilience to cyber attacks \cite{Mylrea2017} have been highlighted as well for centralized energy trading markets. These concerns are summarized in \cite{Thukral2021}.

\begin{figure}
\centering
\includegraphics[width=0.4\textwidth]{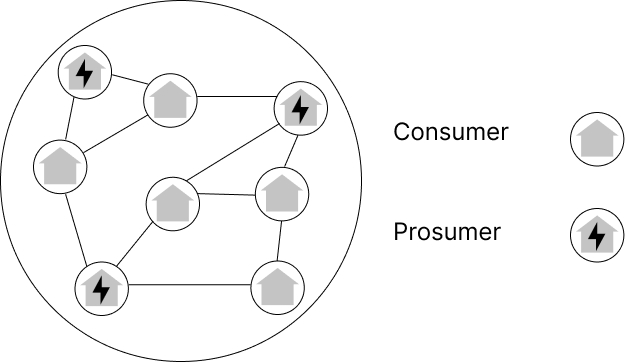}
\caption{Illustration of a microgrid comprising prosumers and consumers.}
\label{microgrid_fig}
\vspace{-1em}
\end{figure}

P2P energy trading is an alternative to centralized markets and it has been explored to address the aforementioned challenges as well as a solution to scalability, privacy control, and transaction speed of the energy network \cite{Murkin2016}.
P2P microgrid systems are a promising implementation of P2P trading. They are considered an efficient way to distribute energy generation between energy sources thereby increasing supply flexibility \cite{Sousa2019}. Microgrids (illustrated in Fig. \ref{microgrid_fig}) are small-scale electricity networks made to supply energy to a small community, e.g. a neighbourhood, a university, or office blocks and can act in grid connected or islanded modes, connecting to other grids through points of common coupling (PCC) \cite{Chowdhury2009}. A typical microgrid contains two system agents, prosumers and consumers, and facilitates P2P energy exchange between these agents through a smart grid infrastructure.

\subsection{Blockchain for P2P Energy Trading}
Blockchain is a key technology for enabling P2P energy trading markets with state-of-the-art in this area reviewed by \cite{Wang2019}. Blockchains not only bring the required decentralisation to the energy markets but also facilitate improvements in privacy, cost, security \cite{Shuaib2018} and transparency \cite{Hwang2017}. Some blockchain technologies are more suited to energy trading than others with the consensus mechanism making some too energy expensive \cite{Nguyen2018},  slow or unscalable \cite{Abdella2021}. The challenges blockchain faces to adoption within the P2P energy trading market are reviewed by \cite{Ahl2019}. 


In \cite{Jabbar2022}, a blockchain-based energy trading system for remote areas is presented. Pi4B hardware is used alongside the Ethereum blockchain and Ethereum smart contracts. Power Ledger \cite{pl} is a blockchain solution built on the Ethereum blockchain (using proof-of-work as consensus) with a long-planned move to the Solana network (using proof-of-stake consensus) and offers low cost, high efficiency energy trading. Pylon \cite{pylon} network takes a different approach by building a blockchain platform from the ground up. It uses a form of proof-of-work consensus on federated nodes and aims to simplify communication between all agents achieving fast, secure and scalable communication. 

Our work aims to present a structural optimization to energy trading systems by offering an alternative topology DLT to blockchain as well as an alternative smart contract technology with the aim of improving cost, speed, and energy demands.

\subsection{IOTA DLT for P2P Energy Trading}
IOTA is a highly scalable DAG-based DLT built for IoT networks and applications. Its role in P2P energy trading has been explored in some previous studies like \cite{Park2019} and \cite{mg}. However, these studies mainly focused on IOTA 1.0 which had an element of centralization in the form of a coordinator and also lacked support for smart contracts and digital assets, thereby making it unsuitable for effectively managing a P2P energy trading system.


With the advent of IOTA 2.0 \cite{IOTA_2}, IOTA has improved on its original goals of decentralization, scalability, and IoT support as well as gaining the ability to fully manage a P2P energy trading system through improved consensus, support for smart contracts, and enhanced security and privacy  \cite{IOTA_2_BRIL}. The ledger layer (Layer-1) is now fully decentralized through the removal of the coordinator. Also, an IOTA smart contract protocol (ISCP) allowing agents to interact over the network securely has been implemented while keeping it scalable. There have also been improvements in the energy efficiency of the system through a lightweight adaptive proof-of-work over conventional proof-of-work which incurs high energy cost. 

A beginner's guide to IOTA and the different types of nodes involved like Wasp nodes (for smart contracts) and GoShimmer nodes (running prototype software that allows to reach consensus) is available at https://iota-beginners-guide.com/category/iota-products/


\subsection{ISCP Protocol}
\begin{figure}
\centering
\includegraphics[width=0.4\textwidth]{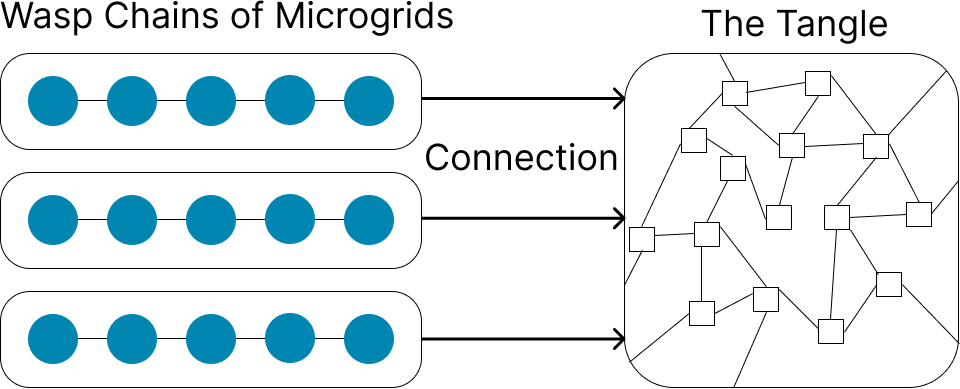}
\caption{Wasp chains executing in parallel, anchoring state of smart contracts to Layer-1 ledger through their connection.}
\label{anchor}
\end{figure}

\begin{figure}
\centering
\includegraphics[width=0.45\textwidth]{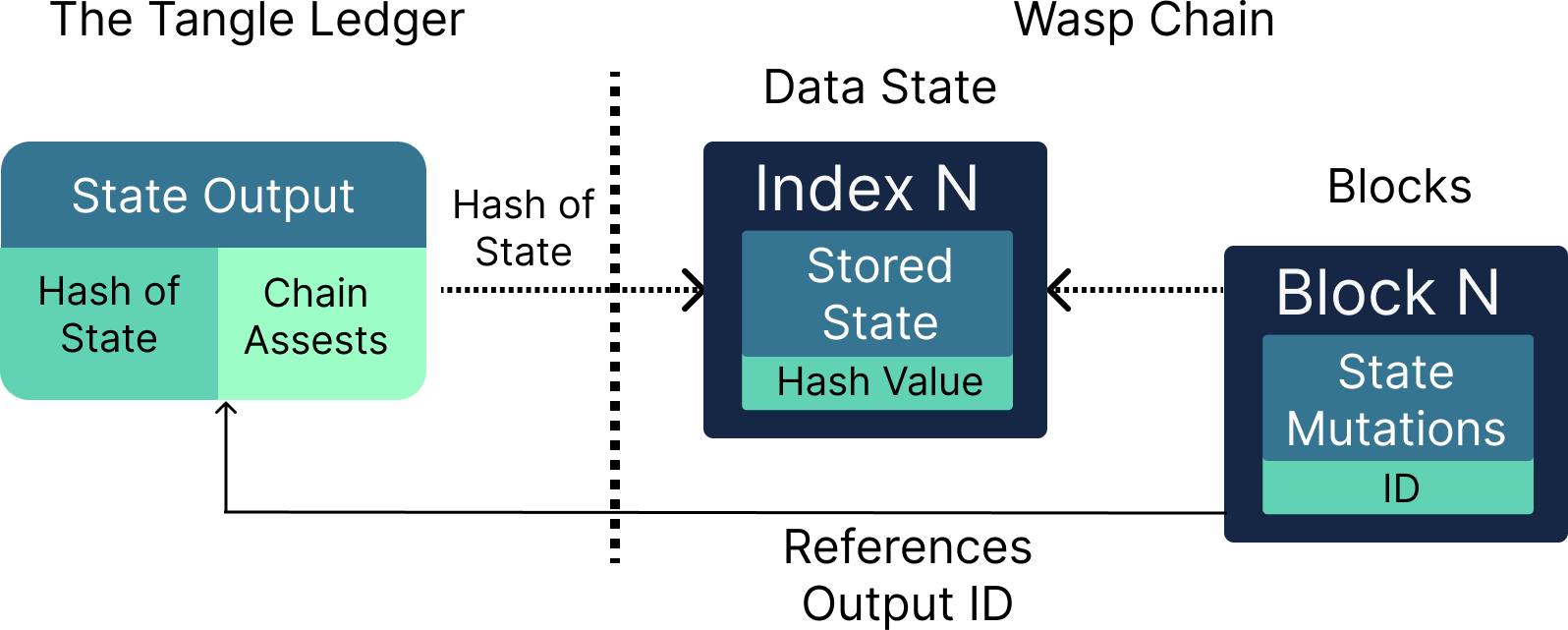}
\caption{Illustration of how a Wasp chain anchors state to the Layer-1 ledger to ensure immutability.}
\label{anchorState}
\vspace{-1em}
\end{figure}

ISCP utilizes sharding to improve  scalibility through splitting up the network into smaller blockchains known as shards, with each shard only validating a subset of transactions \cite{9169448}. IOTA achieves sharding through Wasp chains which are deployed upon a committee of Wasp nodes creating a second layer (Layer-2) platform to run IOTA smart contracts \cite{FoundationIOTAContracts}. Wasp chains execute quickly as they only have to manage and verify the smart contracts deployed upon them. By splitting up the consensus, Wasp chains can run in parallel as shown in Fig. \ref{anchor}. This is a vast improvement on traditional systems which need the entire network to verify every smart contract. ISCP maintains state immutability by anchoring the state of the smart contracts to the Layer-1 ledger ``The Tangle" as shown in Fig. \ref{anchorState} \cite{Foundation}. 

\section{IOTA-based P2P Energy Trading System}
We split the overall system into two parts. First, we describe how the system integrates with IOTA double layered topology to maximize scalability. Then we discuss the market mechanism with a hierarchical routing infrastructure to facilitate P2P energy trading among interconnected microgrids employing a uniform double-auction strategy.

\begin{figure}
\centering
\includegraphics[width=0.4\textwidth]{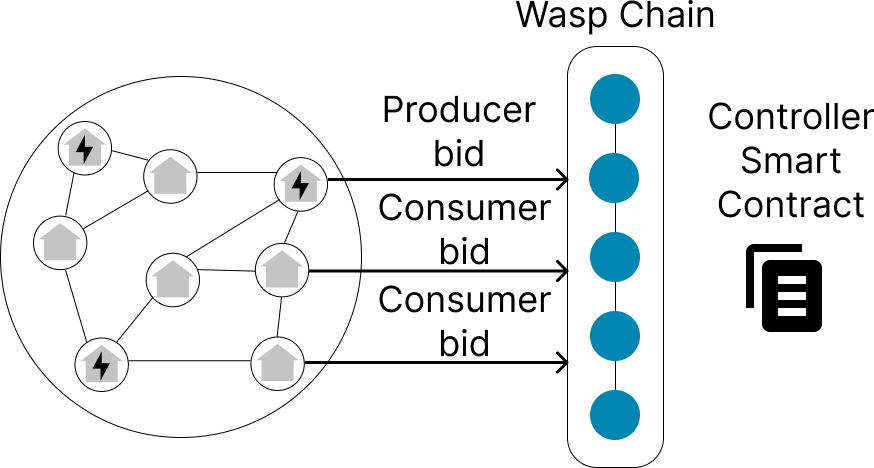}
\caption{Microgrid agents placing bids to the Wasp chain which are handled by the controller smart contract.}
\label{placeBids}
\end{figure}

\subsection{DLT Topology}
Each agent in the microgrid is assumed to run a smart meter. A smart meter is a metering device capable of predicting the energy usage in the next time instance and its role has been explored in some previous studies like \cite{Thakur2019DistributedBlockchains}.

\begin{figure}
\centering
\includegraphics[width=0.4\textwidth]{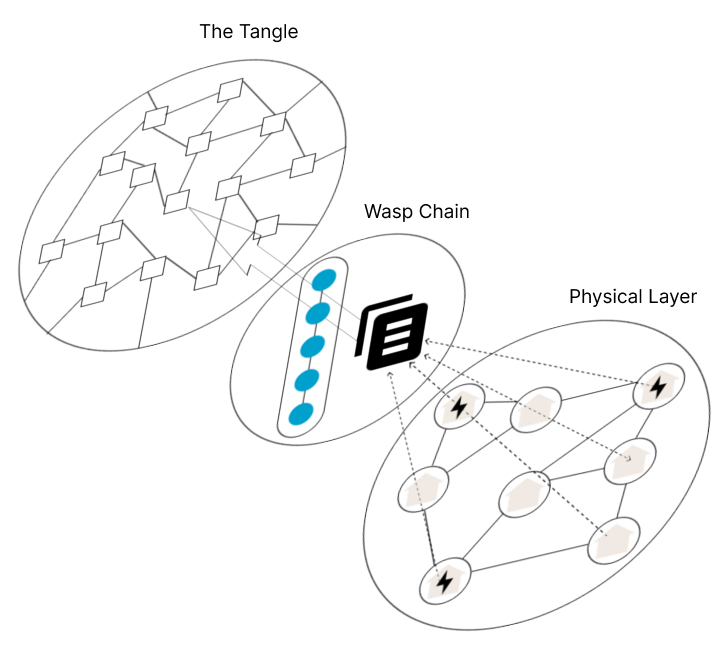}
\caption{CPS representation of IOTA-based P2P energy trading for a single microgrid.}
\label{overview}
\vspace{-1.5em}
\end{figure}

We define \(M = (M_1, M_2, ..., M_n)\) as the set of peers in a microgrid, \(t = (t_1, t_2, ..., t_n)\) as the discrete time instances to trade in, and \(\delta(Mi)\) as the net energy consumption in the following period. 
For our model, we assume each smart meter runs a Wasp node upon it. Since the smart meters are running their individual Wasp nodes they can place bids directly to the smart contracts as shown in Fig. \ref{placeBids} without having interference from other microgrids or traffic on the network as each microgrid is working on a separate Wasp chain. This takes advantage of the Layer-2 framework as the Wasp chains come to price consensus individually before querying connected microgrids and the main grid to handle the surplus/deficit of energy for each bidding round. Separate chains are able to communicate using the Layer-1 ledger as a communication medium. An overview of the trading system implementation, as a manifestation of CPS approach,  is shown in Fig. \ref{overview}.

\subsection{Market Mechanism and Hierarchical Framework} \label{market}
Our trading strategy implements a uniform price double-auction for the market mechanism to match multiple buyers and sellers. This is because it allows the microgrid to act as a single bidding entity when connecting to other grids. The double-auction executes as follows.
\subsubsection{Place Bids}
Bids are placed in the form of buy limit orders (BLOs) submitted by consumers and sell limit orders (SLOs) submitted by producers to conform to the double-auction market mechanism \cite{Pennanen2020EfficientMarkets}. The quantity for the bids \(q_i\) is determined by the energy requirement of the peer \(\delta(Mi)\) defined previously.

A BLO is of the form \((p_i, q_i)\), where \(p_i\) is the maximum price the bidder is willing to pay for a unit of energy. In our model, this is limited by the purchasing price from the main grid, i.e., \(p_i < P_{gb}\), where \(P_{gb}\) denotes the unit price of energy supplied by the main grid. 

An SLO is of the form \((p_i, q_i)\)  where \(p_i\) is the minimum price the bidder is willing to sell a unit of energy. In our model, this is limited by the selling price to the main grid, i.e., \(p_i > P_{gs}\), where \(P_{gs}\) denotes the unit price for sale to the main grid.

After a bid is placed, the double-auction starts and all entities desiring to participate in this auction place their bids. 

\begin{figure}
\centering
\includegraphics[width=0.4\textwidth]{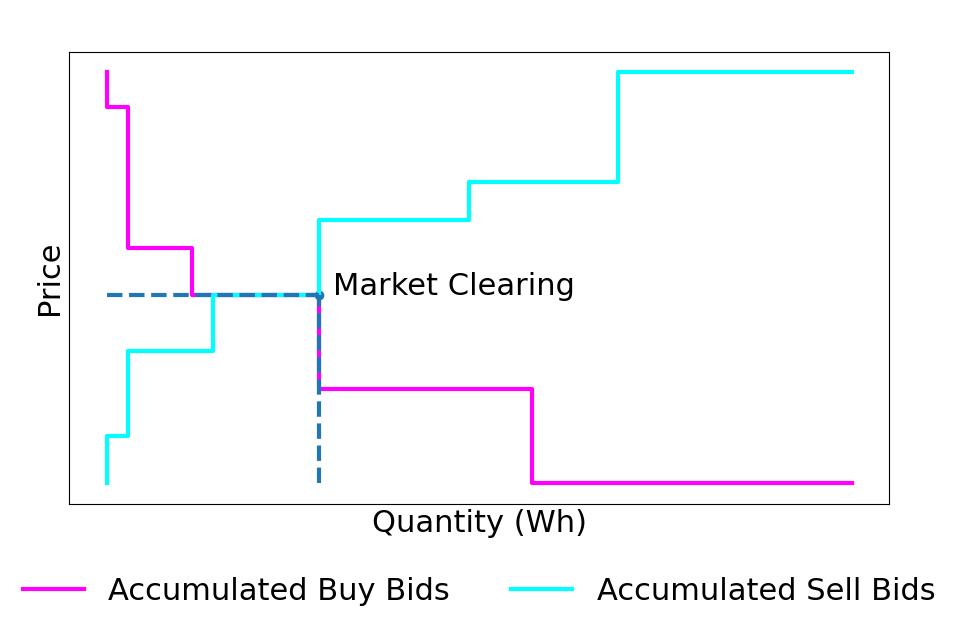}
\caption{Illustration of an example double-auction scenario representing how market clearing price is calculated.}
\label{double}
\vspace{-1em}
\end{figure}

\subsubsection{Price Consensus}
After bids are placed they are ordered in ascending and descending order as shown in Fig. \ref{double} and a market clearing price (MCP) is calculated. This market price is used to update the bids. Due to the time constraints of the system, if consensus cannot be reached in a suitable time we split the difference between the highest consumer bid price \(P_{con}\) and the lowest producer bid \(P_{pro}\) as shown in \eqref{MCP} \cite{9159650}. 

\begin{equation}
    P_{MCP} = \frac{min(P_{con}) + max(P_{pro})}{2}      
    \label{MCP}
\end{equation}

Demand and supply are fixed at the start of each period. Therefore, we need to match all energy within the grid for each period, leaving no wasted energy and adhering to the demand response requirements \cite{RezaDibaj2020APB-DODAM}. Bids are updated in a model of aggressiveness \(r\). We only deal with intramarginal buyers and sellers as we are limited by our reserve prices set at the buy and sell prices to the grid \(P_{gb}\) and \(P_{gs}\). An intramarginal buyer is someone whose limit price is higher than the market equilibrium. An intramarginal seller is someone whose limit price is lower than the market equilibrium. We use an adapted function of \cite{Vytelingum2008StrategicAuctions}  which models the bidding behaviour of intramarginal bidders depending on their aggressiveness. For intramarginal buyers, the bid price \(B_P\) is given by
\begin{equation}
\begin{array}{l}
B_{P} = \begin{cases} 
          c + (MCP - c)(\frac{e^{-2r} - 1}{e^{2} - 1}) & r \in (-1, 0) \\
        MCP + (l - MCP)(1 - \frac{e^{2r} - 1}{e^{2} - 1}) &  r \in (0, 1) 
    \end{cases} 
    \end{array}
 \label{ageq1}   
\end{equation}
where \(MCP\) is the market clearing price, \(c\) is the sell limit, and \(l\) is the buy limit. 
For intramarginal sellers, the bid price \(B_P\) is given by
\begin{equation}
\begin{array}{l}
B_{P} = \begin{cases} 
          MCP + (l - MCP)(1 - \frac{e^{-2r} - 1}{e^{2} - 1}) & r \in (-1, 0) \\
        c + (MCP - c)(\frac{e^{2r} - 1}{e^{2} - 1}) &  r \in (0, 1) 
    \end{cases} 
    \end{array}
 \label{ageq2}   
\end{equation}


\begin{figure}
\centering
\includegraphics[width=0.5\textwidth]{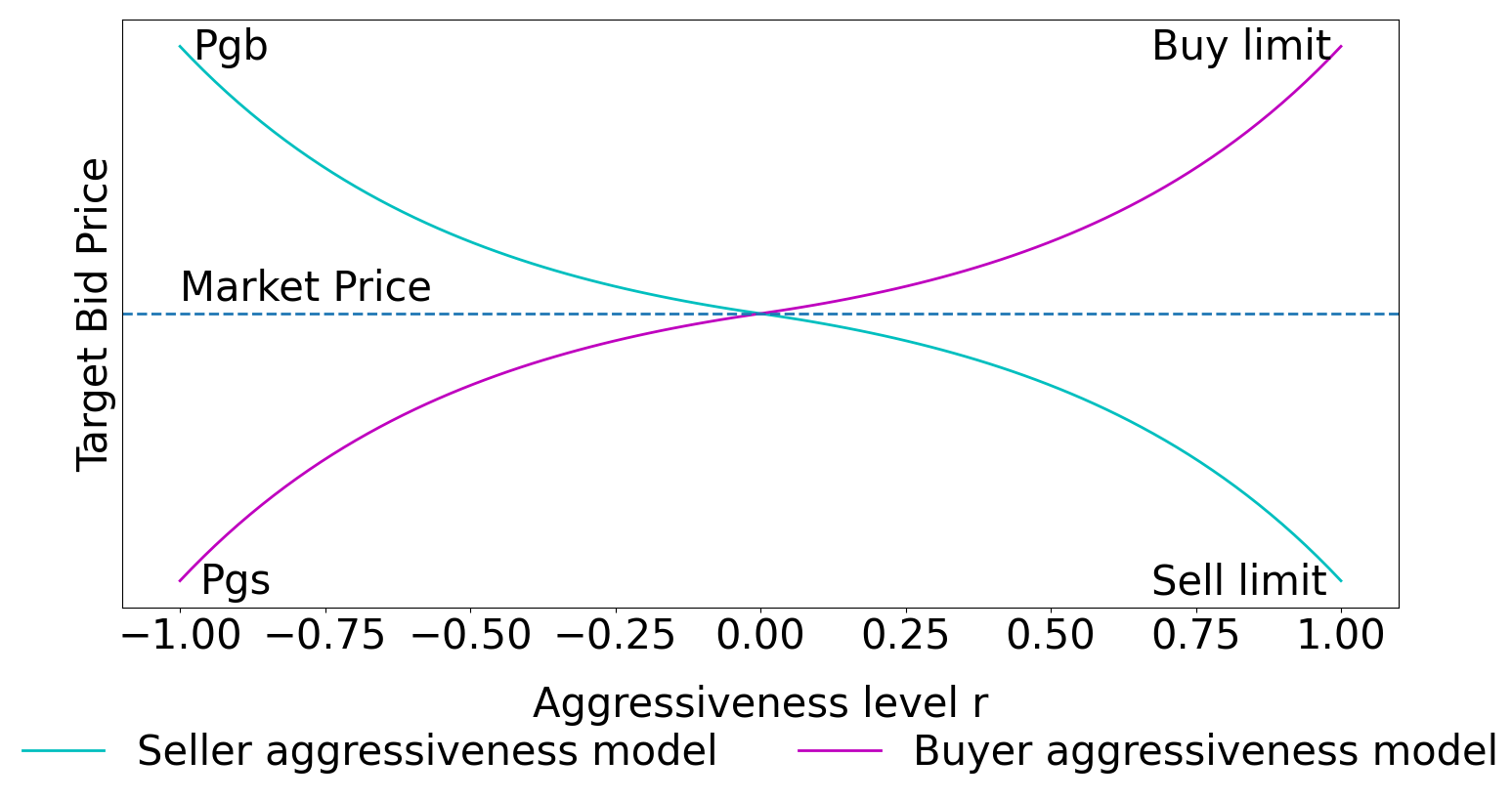}
\caption{Visualization of how bidders place their bids depending on their individual aggressiveness (\(r\)).}
\label{aggressiveness}
\vspace{-1em}
\end{figure}

The bid price is further explained in Fig. \ref{aggressiveness} based on \eqref{ageq1} and \eqref{ageq2}. 
If a bidder is neutral, i.e., \(r = 0\), it aims its next bid at the current market price. However, if a bidder is aggressive, i.e., \(r = 1\), it tries to place its bids to maximize the chances of success. A seller aims its bid at its sell limit price whereas a buyer aims its bid at the maximum buy. If a bidder is completely passive, i.e., \(r = -1\), it aims its bids to maximize profits. A seller tries to place a bid at the maximum ask, i.e., \(P_{gb}\), and a buyer places a bid at the minimum ask, i.e., \(P_{gs}\). 

The double-auction comes to a uniform price consensus determined by how aggressive each bidding entity is. For each update round the bidders update their aggressiveness. The aggressiveness update rules are outlined in Algorithm \ref{rules} and follow the principals of short-term learning for agents acting in double-auctions laid out in \cite{Vytelingum2008StrategicAuctions}. Note that depending on the price of the original bids and the current levels of supply and demand in the system, the entity has to behave in specific ways. If a buyer or seller has freedom, they can choose whether to continue being passive or become more aggressive. However, if the entity does not have freedom due to the level of supply and demand, the entity must be more aggressive in the next update bid round.

\begin{algorithm}
\caption{Learning rules for bidders}\label{rules}
\begin{algorithmic}[1]
\item \textbf{Learning Rules for Buyer i:}
\If{\((Demand < Supply) \And B_{p} < MCP\)}\\
    \text{Buyer must be more aggressive}
\Else \\
\text{Buyer can update bid freely in the market}
\EndIf
\item \textbf{Learning Rules for Seller j:}
\If{\((Demand > Supply) \And B_{P} < MCP\)}\\
\text{Seller must be more aggressive}
\Else \\
\text{Seller can update bid freely in the market}
\EndIf
\end{algorithmic}
\end{algorithm}

\subsubsection{Matching Surplus and Deficit}
If there is surplus demand or supply  not fulfilled in the system, the microgrid enters a similar bidding process among the microgrids it is connected to before defaulting to the main grid if there is still surplus. The cost of the energy from this auction is passed on to each microgrid individually through a uniform price correction. 

\section{Evaluation, Results and Discussion}
\begin{figure}
\centering
\includegraphics[width=0.4\textwidth]{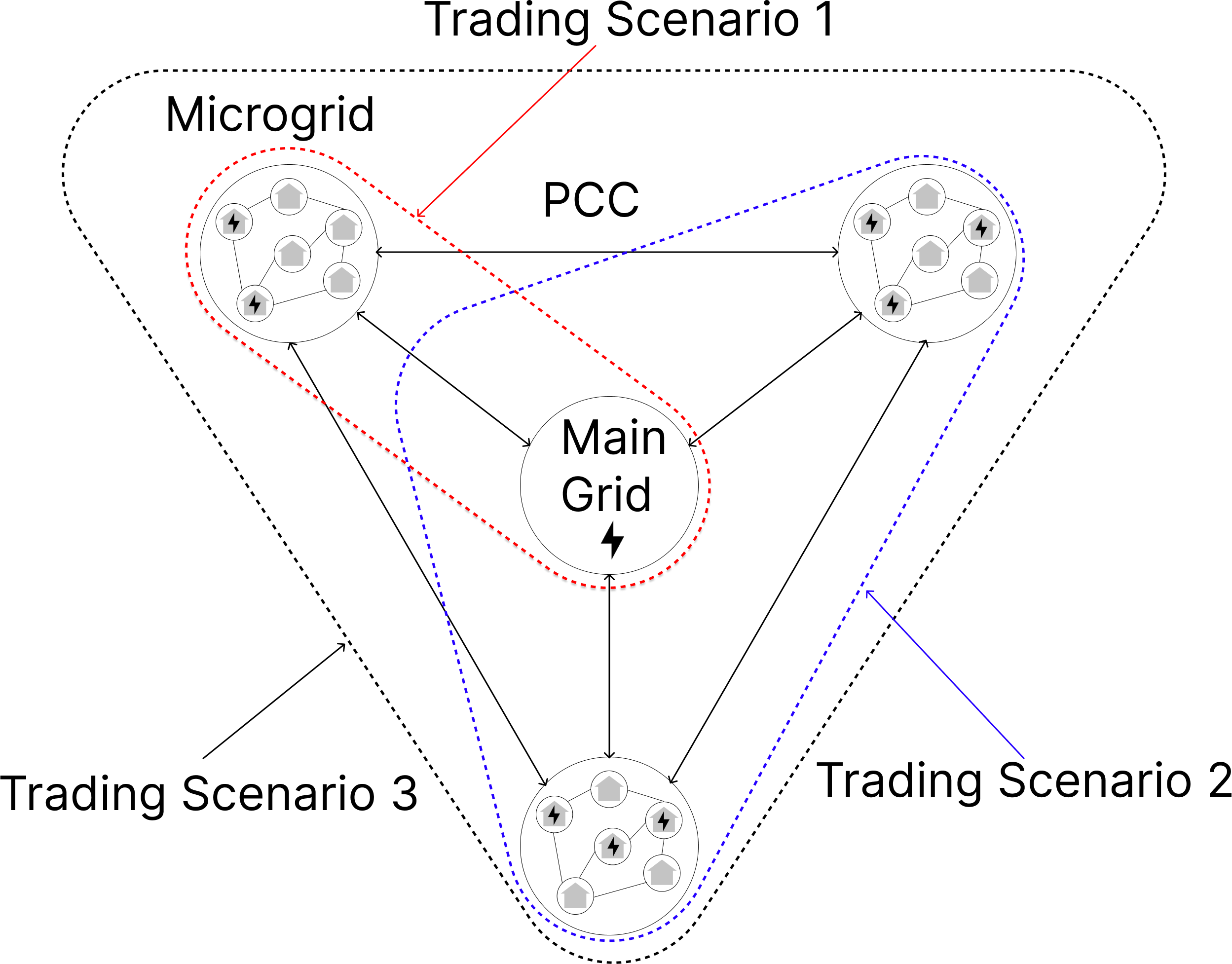}
\caption{Topology of interconnected microgrids (the dotted regions represent microgrids that are being involved in trading via the market mechanism).}
\label{framework}
\end{figure}

The P2P energy trading mechanism is evaluated for both the success as a Layer-2 DLT framework and the benefit of the trading mechanism when applied to an interconnected microgrid system. The system has been implemented on a testbed to prove the functionality over the chosen DLT as well as to obtain metrics from its execution. The smart contracts have been simulated off-ledger to assess the benefit of the market mechanism. In both cases, the system uses real-world data and a python script to model the smart meter prediction capability of production and consumption. We use the script to model the behaviour of each agent in each microgrid when placing bids. The script places the original bids randomly before updating the bid based on Algorithm \ref{rules} and  \eqref{ageq1} and \eqref{ageq2}. The framework of the interconnected microgrids and their trading scenarios is shown in Fig. \ref{framework}.

\begin{figure}
\centering
\includegraphics[width=0.5\textwidth]{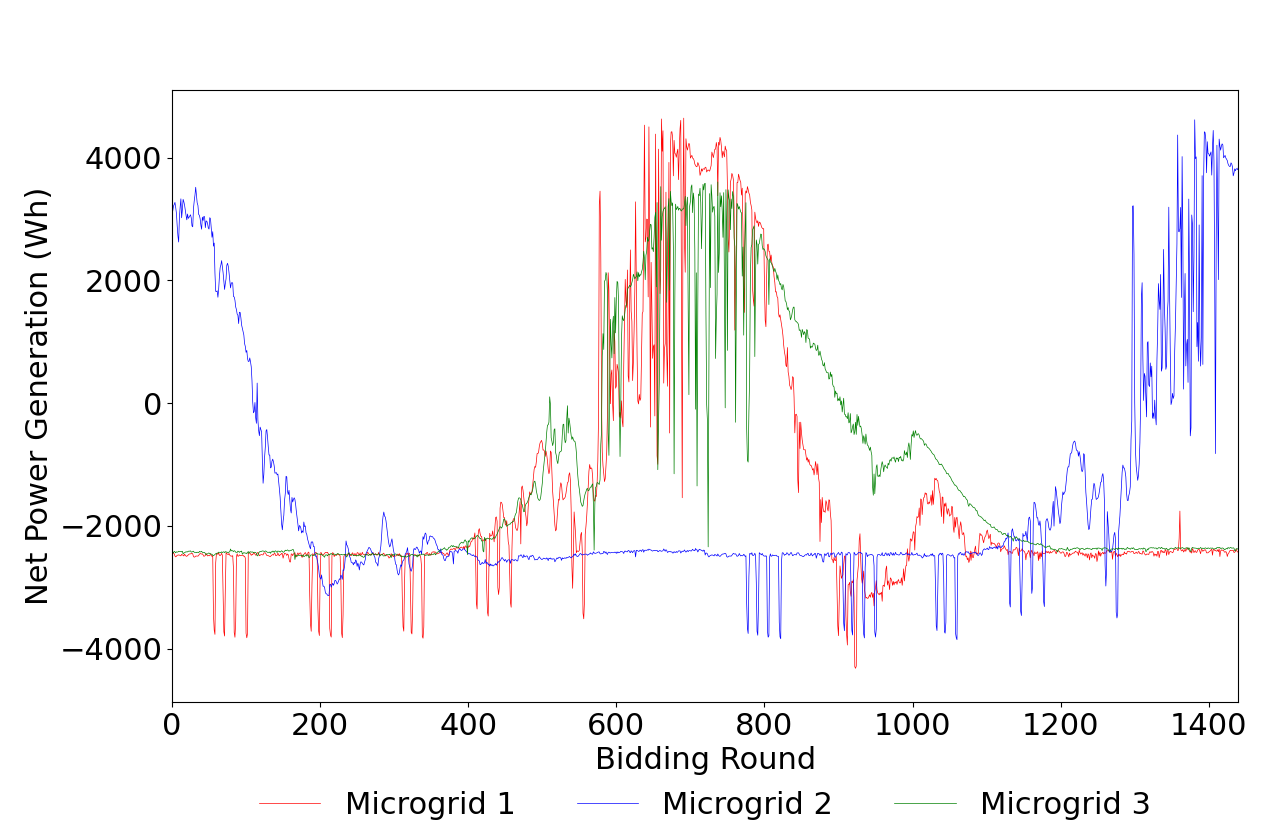}
\caption{Net production and consumption of microgrids.}
\label{NetGen}
\end{figure}

As shown in Fig. \ref{framework}, three microgrids have been modeled to trade and interact with each other and the main grid for a week's period. These microgrids are modeled from \cite{GomesWeekagents}, and each microgrid is made up of five photovoltaic agents logging their production and consumption throughout a week. One of the microgrids has a shifted generation of 12 hours to simulate variable generation of energy from other sources. The net production of energy for each microgrid within the first 24-hour period is represented in Fig. \ref{NetGen} and is representative of the general trends throughout the week.  

\subsection{DLT Topology Results}
\begin{figure}
\centering
\includegraphics[width=0.45\textwidth]{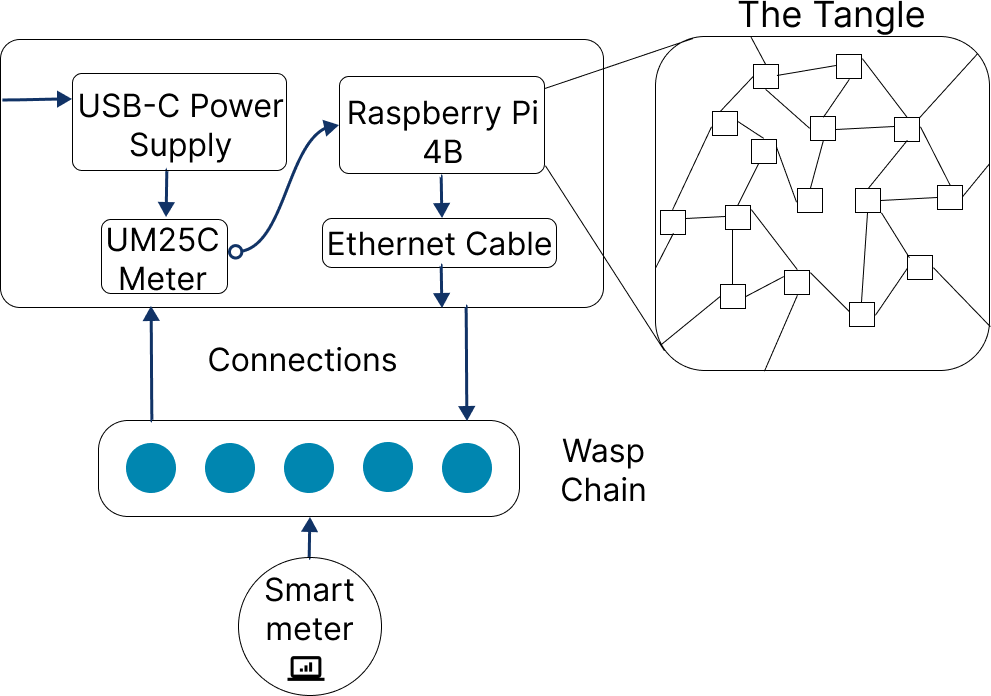}
\caption{TestBed setup for DLT measurements.}
\label{testbed}
\end{figure}

The system is implemented on an experimental testbed as shown in Fig. \ref{testbed}. A Raspberry Pi is used to run a GoShimmer v0.7.7 network handling the Layer-1 ledger. A Laptop is used to model the smart meter, marshalling the data from the data set mentioned above into bids which are sent to the Wasp chain running upon it. The Wasp nodes and the GoShimmer network communicate over the network as our market mechanism executes.

Using the methodology from the IOTA Foundation to measure the energy consumption of a GoShimmer network \cite{LouisHelmer}, we were able to obtain the energy cost per transaction using a Ruideng UM25C energy meter. Through logging the events of the Wasp chain, we averaged the length of time it took for a bid transaction to be confirmed on the DLT. We benchmark against an Ethereum network. With regards to the Ethereum network, the energy cost per transaction and the time taken per transaction are determined through qualitative research\cite{Statista, Ycharts}. The Cost per bid is calculated by deploying an identical uniform double-auction smart contract written in solidity to the Remix IDE, and an average gas fee per bid is calculated.

\begin{table}
    \centering
    \caption{Comparison of Ethereum vs our proposed network}
    \begin{tabular}{ |p{0.2\textwidth}|p{0.1\textwidth}|p{0.1\textwidth}| } 
    \hline
    Metric & IOTA & Ethereum\\
    \hline
    Energy cost per transaction & 3.93nWh & 238.22 kWh \\ 
    \hline
    Monetary cost per bid & 1 IOTA & \(\sim\)\$0.20 \\ 
    \hline
    Time taken per transaction & \(\sim\)12 \text{ seconds} & \(\sim\)15 \text{ seconds} \\ 
    \hline
    Security & Variable & High\\
    \hline
\end{tabular}
    \label{DLT}
\end{table}

At the time of writing this paper 1 IOTA is worth \(\frac{\$0.3303}{1000000}\) 
Table \ref{DLT} conducts a comparison of Ethereum and our network. Ethereum  and our network experience a similar average time per transaction. However, due to the proposed topology, the time per transaction of our system scales better with increased participants. By running the private Wasp chains in parallel, the execution time of each microgrid is independent of others. On the other hand, Ethereum blockchain's transaction time depends on how busy the network is. This causes more unpredictable fluctuations in time, increasing the burden on the capabilities of the smart meter to predict energy consumption and production for variable time periods.
Ethereum gas fees also increase with the traffic on the blockchain, whilst our proposed system has fixed costs. This leads to increased costs as we scale to larger numbers of participants further widening the already huge difference in cost between the two systems. This is also likely to lead to cases where the cost of energy traded in the P2P transactions is less than the gas fee needed to bid for said energy.

Table \ref{DLT} also represents the opportunity cost of the Ethereum system between security and energy cost per transaction. Ethereum outperforms the proposed DLT in security as each transaction in Ethereum is verified by the entire network. However, this means that every transaction has a huge amount of computational power needed to verify it, leading to increased energy cost per transaction. This increases the likelihood that the energy taken to verify transaction is greater than the amount of energy traded, rendering the system economically impractical. On the other hand, our network runs at a fraction of the energy cost per transaction. This is due to the consensus protocol of the DLT. Each transaction only has to do a small amount of work to approve two previous transactions, causing minimal energy consumption. The security of our network is variable as Wasp chains use a committee-based Byzantine Fault Tolerance (BFT) protocol which scales with the number of validators. Committee-based BFT protocol may not be as secure as the Ethereum network but is considered secure when at least two thirds of the network is in consensus.    

\subsection{Market Mechanism Results}
The system was setup off-ledger using the same smart contracts to maintain functionality. Off-ledger simulation was implemented to obtain the results of the mechanism for the entire week period provided by the data set in a timely manner. In this section, we define benefit as the beneficial difference in price agents in the system experience over wholesale alternatives  \(P_{gb}\) and \(P_{gs}\) determined to be fixed at 18.9p/kWh \cite{GoCompareEnergy} and 3.2p/kWh \cite{BristishEarn} respectively for the week period.

\begin{figure}
\centering
\includegraphics[width=0.45\textwidth]{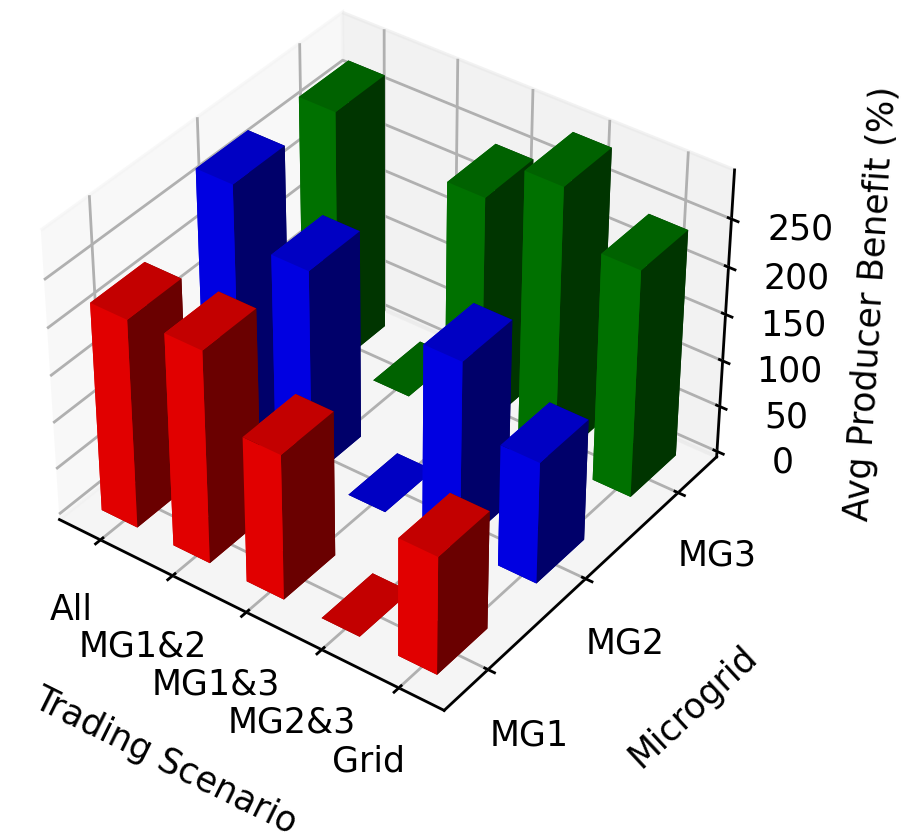}
\caption{The average percentage benefit to producers per Wh traded in different trading scenarios for a week's period.}
\label{averageProd}
\end{figure}

\begin{figure}
\centering
\includegraphics[width=0.45\textwidth]{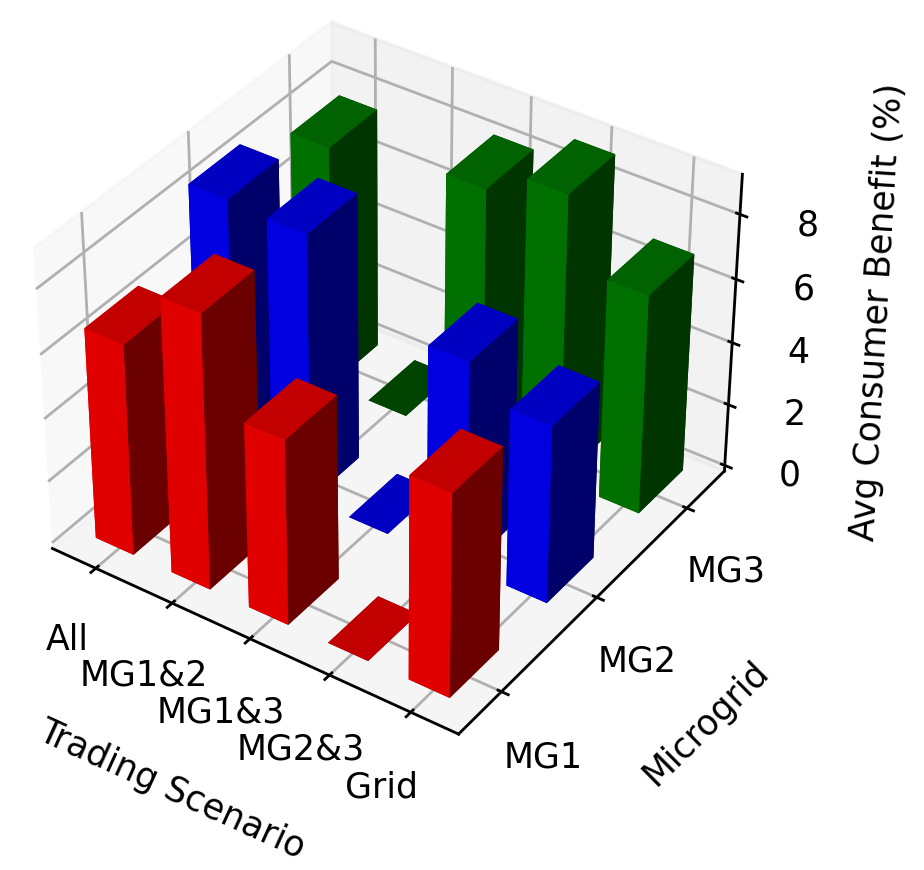}
\caption{The average percentage benefit to consumers per Wh traded in different trading scenarios for a week's period.}
\label{averageCons}
\end{figure}

The average percentage benefit to producers and consumers per watt-hour (Wh) traded in our system are shown in Fig. \ref{averageProd} and Fig. \ref{averageCons}, respectively. These graphs show the baseline benefit over wholesale options in their ``Grid" trading scenario. This metric follows trading scenario 1 shown in Fig. \ref{framework} where each microgrid individually trades energy amongst its peers and the main grid, without querying other microgrids to fulfil excess demand/supply. These graphs also show the benefit of the system in cases of trading scenario 2 with multiple combinations of microgrids trading with each other and the main grid. For metrics in trading scenario 2, only the benefit experienced by those involved in the trading scenario is logged. Finally ``All" represents trading scenario 3 in which all entities participate in the system, trading with each other and the main grid.

In Fig. \ref{averageProd} and Fig. \ref{averageCons} the benefit to producers is greater than the benefit to consumers in every trading scenario. This is due to the level of demand and supply in the system and the update rules for bidders in Algorithm \ref{rules}. From Fig. \ref{NetGen}, we determine that for a majority of the bidding rounds, demand far exceeds supply. This affects the update rules of Algorithm \ref{rules} to make the buyers more aggressive (r trends to 1) and increases the cost of energy. Since energy prices are higher for longer, the producers experience a more significant benefit. 

The effect of demand and supply on our market mechanism is also represented in trading, as MG1 and MG3 experience maximum producer benefit when trading in scenario 2 with MG2. This is because MG2 has a high level of consumption while MG1 and MG3 have high levels of production. Therefore, when the entities are trading with each other they are more likely to experience higher prices, as buyer aggressiveness trends to 1. We see a much decreased benefit to MG1 and MG3 when in trading scenario 2 with each other, as although there is a larger overall increase in demand compared to supply, this is not as significant as when they are trading with MG2. Therefore buyers are less aggressive in their updated pricing behaviours. This logic works both ways,  with consumers experiencing their most significant benefit when there are higher levels of production. Interestingly, this is also experienced when MG1 and MG3 are trading with MG2. Since we are limited by \(P_{gb}\) and \(P{gs}\) the lower prices experienced by MG1 and MG3 when they are trading with each other are not low enough for long enough to outperform the consistently lower than wholesale prices experienced when trading with MG2 as suggested in Fig. \ref{NetGen}. This suggests that further distributing energy generation among multiple sources to supply energy consistently throughout the day is more beneficial to both producers and consumers than short periods of high volume concurrent energy generation. 

\begin{figure}
\centering
\includegraphics[width=0.48\textwidth]{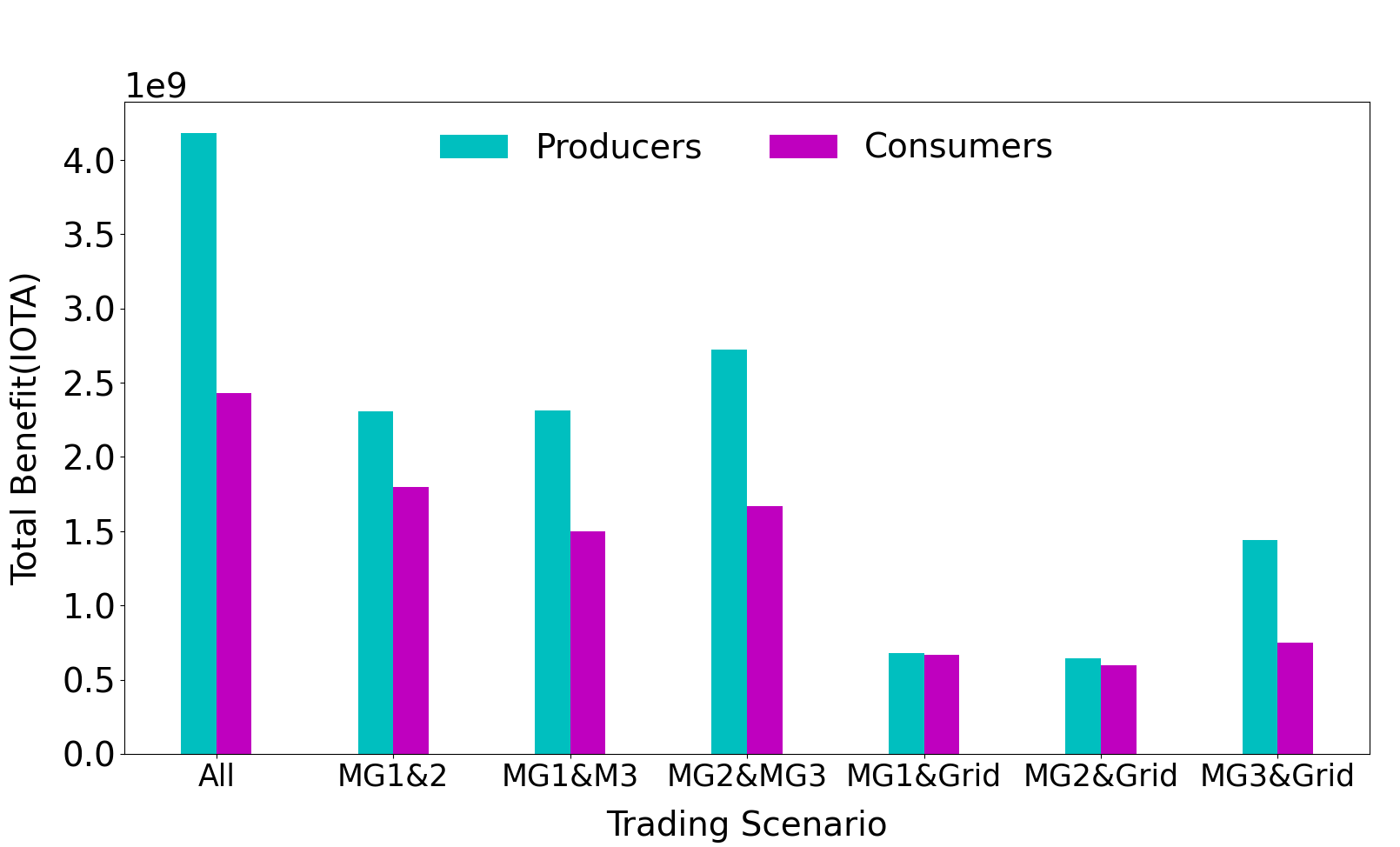}
\caption{Total benefits to the agents involved directly in the trading scenario for both producers and consumers.
}
\label{totalBenefit}
\end{figure}

\begin{figure}
\centering
\includegraphics[width=0.48\textwidth]{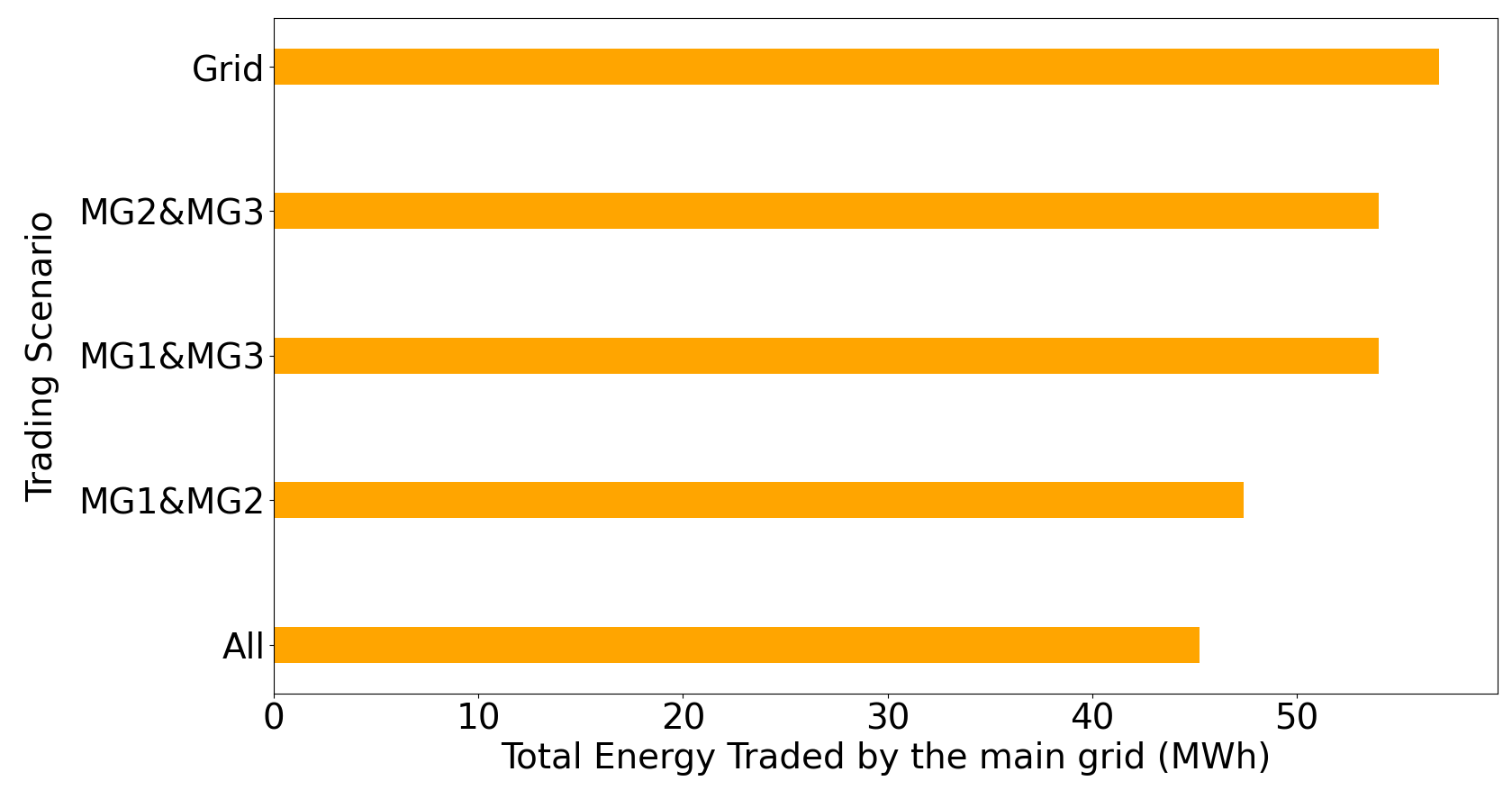}
\caption{Total energy traded with the main grid in different trading scenarios.}
\label{trade}
\end{figure}

Fig. \ref{averageProd} shows that producers experience their maximum benefit in trading scenario 3 where all microgrids are connected to and trading with each other. Fig. \ref{averageCons} also shows a marked improvement over trading scenario 1 for consumers in this trading scenario. This is further emphasized in Fig. \ref{totalBenefit} showing the total benefit to entities involved in the trading scenario 3 outperforms the benefits of other trading scenarios implying that benefit is maximized as we increase the number of participants in the network. However, the most significant benefit of the ``All" trading scenario is demonstrated in Fig. \ref{trade} which represents the involvement of the main grid in different trading scenarios. If a microgrid is not involved in the trading scenario it is assumed to be trading in scenario 1 with the main grid and the energy it trades with the main grid is included in the total.

Fig. \ref{trade} shows that the ``All'' trading scenario requires the least amount of energy to be traded from the main grid. This represents that the more microgrids that interact and connect in our P2P trading mechanism the less reliant we are on the main grid, reducing risks of single point-of-failure. This is because every microgrid is trading with each other, and therefore, there is an increased likelihood that another microgrid can make up for the surplus demand/supply of energy in that time period instead of the main grid, increasing the efficiency of our energy distribution within the network. This is a significant factor of the greater benefits represented in Fig. \ref{totalBenefit} as agents take advantage of the more beneficial prices of energy produced in the system over the main grid.


\section{Concluding Remarks}
In this paper, a DLT-based P2P energy trading system aimed at exploiting the benefits of IOTA Tangle 2.0 and ISCP was shown to only be a fraction of the cost of traditional blockchains for both energy and monetary costs. It achieved this whilst maintaining the inherent benefits provided by DLT and smart contracts of transparency, security and privacy in a completely decentralized manner. The implemented system is also proven to have a significant monetary benefit to participants over wholesale options, with benefits to both prosumers and consumers. The microgrids experienced an average of 166.41\% benefit to producers and a 6.3\% benefit to consumers in the grid connected scenario per Wh traded. The hierarchical routing infrastructure of interconnected microgrids and the main grid fulfilled the excess energy needs of the system. Additionally, the structure was shown to reduce the overall reliance on the main grid as more entities traded together, increasing benefits to agents in the system. There was shown to be 20.6\% less energy traded with the main grid when comparing the ``All" trading of microgrids in trading scenario 3 over the ``Grid" scenario where each microgrid was trading energy in trading scenario 1, implying the energy generated within the system was distributed more efficiently. 

The work presented in this paper can be further extended in the following ways.
\begin{itemize}
    \item \textbf{Energy store:} An energy store helps smooth the intermittent nature of DER energy generation, allowing agents to store excess energy increasing efficiency. These could take the form of electric vehicles  with energy storage and transport capabilities. 
    \item \textbf{Stardust network:} IOTA continues to develop its ISCP, moving towards the stardust network, a production-ready implementation of the multi-layered infrastructure. It would be important to analyze how the increased  efficiency of the system benefits leads to a more optimized framework.
\end{itemize}

\bibliographystyle{IEEEtran}
\bibliography{references.bib}

\begin{thebibliography}{10}
\providecommand{\url}[1]{#1}
\csname url@samestyle\endcsname
\providecommand{\newblock}{\relax}
\providecommand{\bibinfo}[2]{#2}
\providecommand{\BIBentrySTDinterwordspacing}{\spaceskip=0pt\relax}
\providecommand{\BIBentryALTinterwordstretchfactor}{4}
\providecommand{\BIBentryALTinterwordspacing}{\spaceskip=\fontdimen2\font plus
\BIBentryALTinterwordstretchfactor\fontdimen3\font minus
  \fontdimen4\font\relax}
\providecommand{\BIBforeignlanguage}[2]{{%
\expandafter\ifx\csname l@#1\endcsname\relax
\typeout{** WARNING: IEEEtran.bst: No hyphenation pattern has been}%
\typeout{** loaded for the language `#1'. Using the pattern for}%
\typeout{** the default language instead.}%
\else
\language=\csname l@#1\endcsname
\fi
#2}}
\providecommand{\BIBdecl}{\relax}
\BIBdecl

\bibitem{ZHANG20181}
\BIBentryALTinterwordspacing
C.~Zhang \emph{et~al.}, ``{Peer-to-Peer energy trading in a Microgrid},''
  \emph{Applied Energy}, vol. 220, pp. 1--12, 2018. [Online]. Available:
  \url{https://www.sciencedirect.com/science/article/pii/S0306261918303398}
\BIBentrySTDinterwordspacing

\bibitem{Arumugam2022}
R.~Arumugam and T.~Subbaiyan, ``A review of dynamic pricing and peer-to-peer
  energy trading in smart cities with emphasize on electric vehicles,'' pp.
  1--6, 6 2022.

\bibitem{Tushar2020}
Tushar \emph{et~al.}, ``{Peer-to-Peer Trading in Electricity Networks: An
  Overview},'' \emph{IEEE Transactions on Smart Grid}, vol.~11, no.~4, pp.
  3185--3200, 2020.

\bibitem{Thukral2021}
\BIBentryALTinterwordspacing
M.~K. Thukral, ``Emergence of blockchain-technology application in peer-to-peer
  electrical-energy trading: a review,'' \emph{Clean Energy}, vol.~5, pp.
  104--123, 3 2021. [Online]. Available:
  \url{https://academic.oup.com/ce/article/5/1/104/6166927}
\BIBentrySTDinterwordspacing

\bibitem{mg}
J.~Garcia-Hernandez, L.~G. Marin-Collazos, G.~Jimenez-Estevez, and
  P.~Mendoza-Araya, ``Distributed ledger technologies based microgrid energy
  management using iota tangle,'' \emph{2021 IEEE CHILEAN Conference on
  Electrical, Electronics Engineering, Information and Communication
  Technologies, CHILECON 2021}, 2021.

\bibitem{Aitzhan2018}
N.~Z. Aitzhan and D.~Svetinovic, ``Security and privacy in decentralized energy
  trading through multi-signatures, blockchain and anonymous messaging
  streams,'' \emph{IEEE Transactions on Dependable and Secure Computing},
  vol.~15, pp. 840--852, 9 2018.

\bibitem{Gregoratti2015}
D.~Gregoratti and J.~Matamoros, ``Distributed energy trading: The
  multiple-microgrid case,'' \emph{IEEE Transactions on Industrial
  Electronics}, vol.~62, pp. 2551--2559, 2015.

\bibitem{Guan2019}
\BIBentryALTinterwordspacing
Z.~Guan \emph{et~al.}, ``Effect: an efficient flexible privacy-preserving data
  aggregation scheme with authentication in smart grid,'' \emph{Science China
  Information Sciences 2019 62:3}, vol.~62, pp. 1--14, 1 2019. [Online].
  Available: \url{https://link.springer.com/article/10.1007/s11432-018-9451-y}
\BIBentrySTDinterwordspacing

\bibitem{Ettlin2018}
A.~Ettlin, ``Dynamic modeling of peer-to-peer power market making,''
  \emph{International Conference on the European Energy Market, EEM}, vol.
  2018-June, 9 2018.

\bibitem{Mylrea2017}
M.~Mylrea and S.~N.~G. Gourisetti, ``Blockchain for smart grid resilience:
  Exchanging distributed energy at speed, scale and security,''
  \emph{Proceedings - 2017 Resilience Week, RWS 2017}, pp. 18--23, 10 2017.

\bibitem{Murkin2016}
\BIBentryALTinterwordspacing
J.~Murkin, R.~Chitchyan, and A.~Byrne, ``Enabling peer-to-peer electricity
  trading,'' pp. 234--235, 8 2016. [Online]. Available:
  \url{https://www.atlantis-press.com/proceedings/ict4s-16/25860390}
\BIBentrySTDinterwordspacing

\bibitem{Sousa2019}
T.~Sousa \emph{et~al.}, ``Peer-to-peer and community-based markets: A
  comprehensive review,'' \emph{Renewable and Sustainable Energy Reviews}, vol.
  104, pp. 367--378, 4 2019.

\bibitem{Chowdhury2009}
\BIBentryALTinterwordspacing
S.~Chowdhury and P.~Crossley, \emph{{Microgrids and Active Distribution
  Networks}}.\hskip 1em plus 0.5em minus 0.4em\relax Institution of Engineering
  and Technology, jan 2009. [Online]. Available:
  \url{https://digital-library.theiet.org/content/books/po/pbrn006e}
\BIBentrySTDinterwordspacing

\bibitem{Wang2019}
\BIBentryALTinterwordspacing
N.~Wang \emph{et~al.}, ``When energy trading meets blockchain in electrical
  power system: The state of the art,'' \emph{Applied Sciences 2019, Vol. 9,
  Page 1561}, vol.~9, p. 1561, 4 2019. [Online]. Available:
  \url{https://www.mdpi.com/2076-3417/9/8/1561/htm
  https://www.mdpi.com/2076-3417/9/8/1561}
\BIBentrySTDinterwordspacing

\bibitem{Shuaib2018}
K.~Shuaib, J.~A. Abdella, F.~Sallabi, and M.~Abdel-Hafez, ``Using blockchains
  to secure distributed energy exchange,'' \emph{2018 5th International
  Conference on Control, Decision and Information Technologies, CoDIT 2018},
  pp. 622--627, 6 2018.

\bibitem{Hwang2017}
J.~Hwang \emph{et~al.}, ``Energy prosumer business model using blockchain
  system to ensure transparency and safety,'' \emph{Energy Procedia}, vol. 141,
  pp. 194--198, 12 2017.

\bibitem{Nguyen2018}
V.~H. Nguyen, Y.~Besanger, Q.~T. Tran, and M.~T. Le, ``On the applicability of
  distributed ledger architectures to peer-to-peer energy trading framework,''
  \emph{IEEE International Conference on Environment and Electrical Engineering
  and IEEE Industrial and Commercial Power Systems Europe}, 10 2018.

\bibitem{Abdella2021}
J.~Abdella \emph{et~al.}, ``An architecture and performance evaluation of
  blockchain-based peer-to-peer energy trading,'' \emph{IEEE Transactions on
  Smart Grid}, vol.~12, pp. 3364--3378, 7 2021.

\bibitem{Ahl2019}
A.~Ahl, M.~Yarime, K.~Tanaka, and D.~Sagawa, ``Review of blockchain-based
  distributed energy: Implications for institutional development,''
  \emph{Renewable and Sustainable Energy Reviews}, vol. 107, pp. 200--211, 6
  2019.

\bibitem{Jabbar2022}
M.~Jabbar \emph{et~al.}, ``A low-cost, open-source peer-to-peer energy trading
  system for a remote community using the internet-of-things, blockchain, and
  hypertext transfer protocol,'' \emph{Energies 2022, Vol. 15, Page 4862},
  vol.~15, p. 4862, 7 2022.

\bibitem{pl}
\BIBentryALTinterwordspacing
``Blockchain technology.'' [Online]. Available:
  \url{https://www.powerledger.io/blockchain-technology}
\BIBentrySTDinterwordspacing

\bibitem{pylon}
\BIBentryALTinterwordspacing
``The energy blockchain platform. ® https://pylon-network.org.'' [Online].
  Available: \url{https://pylon-network.org/}
\BIBentrySTDinterwordspacing

\bibitem{Park2019}
\BIBentryALTinterwordspacing
J.~Park, R.~Chitchyan, A.~Angelopoulou, and J.~Murkin, ``A block-free
  distributed ledger for p2p energy trading: Case with iota?'' \emph{Lecture
  Notes in Computer Science (including subseries Lecture Notes in Artificial
  Intelligence and Lecture Notes in Bioinformatics}, pp. 111--125, 2019.
  [Online]. Available:
  \url{http://www.bristol.ac.uk/red/research-policy/pure/user-guides/ebr-terms/}
\BIBentrySTDinterwordspacing

\bibitem{IOTA_2}
\BIBentryALTinterwordspacing
``{Towards Full Decentralization with IOTA 2.0}.'' [Online]. Available:
  \url{https://blog.iota.org/path-towards-full-decentralization-with-iota-2-0/}
\BIBentrySTDinterwordspacing

\bibitem{IOTA_2_BRIL}
\BIBentryALTinterwordspacing
N.~Sealey, A.~Aijaz, and B.~Holden, ``{IOTA Tangle 2.0: Toward a Scalable,
  Decentralized, Smart, and Autonomous IoT Ecosystem},'' in \emph{IEEE
  SmartNets}, 2022. [Online]. Available:
  \url{https://arxiv.org/pdf/2209.04959.pdf}
\BIBentrySTDinterwordspacing

\bibitem{9169448}
S.~Kantesariya and D.~Goswami, ``Determining optimal shard size in a
  hierarchical blockchain architecture,'' in \emph{2020 IEEE International
  Conference on Blockchain and Cryptocurrency (ICBC)}, 2020, pp. 1--3.

\bibitem{FoundationIOTAContracts}
\BIBentryALTinterwordspacing
I.~Foundation, ``{IOTA Smart Contracts}.'' [Online]. Available:
  \url{https://wiki.iota.org/smart-contracts/overview}
\BIBentrySTDinterwordspacing

\bibitem{Foundation}
\BIBentryALTinterwordspacing
{IOTA Foundation}, ``{State}.'' [Online]. Available:
  \url{https://wiki.iota.org/smart-contracts/guide/core_concepts/states}
\BIBentrySTDinterwordspacing

\bibitem{Thakur2019DistributedBlockchains}
Thakur \emph{et~al.}, ``{Distributed double auction for peer to peer energy
  trade using blockchains},'' 2019.

\bibitem{Pennanen2020EfficientMarkets}
T.~Pennanen, ``{Efficient Allocations in Double Auction Markets},'' 2020.

\bibitem{9159650}
Z.~Zhao \emph{et~al.}, ``Energy transaction for multi-microgrids and internal
  microgrid based on blockchain,'' \emph{IEEE Access}, vol.~8, pp.
  144\,362--144\,372, 2020.

\bibitem{RezaDibaj2020APB-DODAM}
R.~Dibaj \emph{et~al.}, ``{A cloud priority-based dynamic online double auction
  mechanism (PB-DODAM)},'' \emph{Journal of Cloud Computing}, vol.~9, no.~1,
  2020.

\bibitem{Vytelingum2008StrategicAuctions}
P.~Vytelingum, D.~Cliff, and N.~R. Jennings, ``{Strategic bidding in continuous
  double auctions},'' pp. 1700--1729, 2008.

\bibitem{GomesWeekagents}
\BIBentryALTinterwordspacing
L.~Gomes and Z.~Vale, ``{week monitorization data of a microgrid with five
  agents}.'' [Online]. Available:
  \url{https://zenodo.org/record/3371339#.Yp8ZxTnMJH5}
\BIBentrySTDinterwordspacing

\bibitem{LouisHelmer}
\BIBentryALTinterwordspacing
L.~Helmer and A.~Penzkofer, ``{Report on the energy consumption of the
  GoShimmer network}.'' [Online]. Available:
  \url{https://files.iota.org/papers}
\BIBentrySTDinterwordspacing

\bibitem{Statista}
\BIBentryALTinterwordspacing
Statista, ``{Ethereum average energy consumption per transaction compared to
  that of VISA as of January 10, 2022}.'' [Online]. Available:
  \url{https://www.statista.com/statistics/1265891/ethereum-energy-consumption-transaction-comparison-visa/}
\BIBentrySTDinterwordspacing

\bibitem{Ycharts}
\BIBentryALTinterwordspacing
Ycharts, ``{Ethereum Average Block Time}.'' [Online]. Available:
  \url{https://ycharts.com/indicators/ethereum\_average\_block\_time}
\BIBentrySTDinterwordspacing

\bibitem{GoCompareEnergy}
\BIBentryALTinterwordspacing
``{GoCompare Energy}.'' [Online]. Available:
  \url{https://www.gocompare.com/gas-and-electricity/guide/energy-per-kwh/}
\BIBentrySTDinterwordspacing

\bibitem{BristishEarn}
\BIBentryALTinterwordspacing
``{Bristish Gas Export and Earn}.'' [Online]. Available:
  \url{https://www.britishgas.co.uk/business/help-and-support/billing-and-payments/smart-export-guarantee}
\BIBentrySTDinterwordspacing

\end{thebibliography}

\end{document}